\documentclass[iop,apj]{emulateapj}   % makes it look like the ApJ format

\pdfoutput=1

\usepackage{graphicx}
\usepackage{epsfig}
\usepackage{amsmath}
\usepackage{subfigure} 
\usepackage{hyperref}
\usepackage{amssymb}
\usepackage{apjfonts}
\usepackage{latexsym}
\usepackage{color}
\usepackage[normalem]{ulem}
\usepackage[usenames,dvipsnames,svgnames,table]{xcolor}
\usepackage{hyperref}
\usepackage[all]{hypcap}    %for going to the top of an image when a figure reference is clicked
\hypersetup{
    colorlinks,
    citecolor=blue,
    filecolor=blue,
    linkcolor=blue,
    urlcolor=blue
}

\def\kms{\ifmmode{~{\rm km~s^{-1}}}\else{~km s$^{-1}$}\fi}
\def\cc{\ifmmode{~{\rm cm^{-3}}}\else{~cm$^{-3}$}\fi}
\def\fesc{\ifmmode{{f_{esc}}}\else{$f_{\rm esc}$}\fi}
\def\fstar{\ifmmode{{f_\star}}\else{$f_\star$}\fi}
\def\lsim{\lower0.3em\hbox{$\,\buildrel <\over\sim\,$}}
\def\gsim{\lower0.3em\hbox{$\,\buildrel >\over\sim\,$}}

\def\enzo{{\sc Enzo}}
\def\moray{{\sc Moray}}
\def\yt{{\sc yt}}
\def\Ms{\ifmmode{~{\rm M_\odot}}\else{M$_\odot$}\fi}
\def\Zs{\ifmmode{~{\rm Z_\odot}}\else{Z$_\odot$}\fi}
\def\h2{H$_2$}

\def\hh{H$_2$}

\def\apj{ApJ}

\def\apjl{ApJL}
\def\apjs{ApJS}
\def\mnras{MNRAS}

\def\aap{A\&A}

\newcommand\unit[1]{\; \textrm{#1}}

\shorttitle{Late Pop III Star Formation} 
\shortauthors{H. Xu et al.}

\begin{document}\title{Late Pop III Star Formation During the Epoch of Reionization: Results from the {\em Renaissance Simulations}}

\author{
  Hao Xu\altaffilmark{1}, 
  Michael L. Norman\altaffilmark{1,2},
  Brian W. O'Shea\altaffilmark{3,4,5,6}, and
  John H. Wise\altaffilmark{7}}

\affil{$^{1}${San Diego Supercomputer Center, University of California, San Diego, 9500 Gilman
    Drive, La Jolla, CA 92093;
    \href{mailto:hxu@ucsd.edu}{hxu@ucsd.edu},
    \href{mailto:mlnorman@ucsd.edu}{mlnorman@ucsd.edu}}}

\affil{$^{2}${CASS, University of California, San Diego, 9500 Gilman
    Drive, La Jolla, CA 92093;}}

\affil{$^{3}${Department of Physics and Astronomy, Michigan State
    University, East Lansing, MI 48824, USA;
    \href{mailto:oshea@msu.edu}{oshea@msu.edu}}}

\affil{$^{4}${Department of Computational Mathematics, Science and Engineering, Michigan State University, East
    Lansing, MI 48824, USA}}

\affil{$^{5}${National Superconducting Cyclotron Laboratory, Michigan State
    University, East Lansing, MI 48824, USA}}

\affil{$^{6}${JINA: Joint Institute for Nuclear Astrophysics}}

\affil{$^{7}${Center for Relativistic Astrophysics, School of Physics,
    Georgia Institute of Technology, 837 State Street, Atlanta, GA
    30332; \href{mailto:jwise@gatech.edu}{jwise@gatech.edu}}}

\label{firstpage}

\begin{abstract}
We present results on the formation of Pop III stars 
at redshift 7.6 from the \textit{Renaissance Simulations}, a suite of extremely
high-resolution and physics-rich radiation transport hydrodynamics cosmological 
adaptive-mesh refinement simulations of high redshift galaxy formation performed on the Blue Waters supercomputer.
In a survey volume of about 220 comoving Mpc$^3$, we found 14 Pop III galaxies with
recent star formation. The surprisingly late formation of Pop III stars
is possible due to two factors: (i) the metal enrichment process is local and slow, leaving plenty of pristine gas to exist in the vast volume; and (ii) strong Lyman-Werner radiation 
from vigorous metal-enriched star formation in early galaxies suppresses Pop III formation in ("not so") small primordial halos with
mass less than $\sim$ 3 $\times$ 10$^7$ M$_\odot$. We quantify the properties of
these Pop III galaxies and their Pop III star formation environments. We look for 
analogues to the recently discovered luminous Ly $\alpha$ emitter CR7 \citep{Sobral15}, 
which has been interpreted as a Pop III star cluster within or near a metal-enriched star forming galaxy. 
We find and discuss a system similar to this in some respects, however the Pop III star cluster is far less  massive and luminous than CR7 is inferred to be.  

\end{abstract}


\section{Introduction}
\label{sec:introduction}

Although Population III stars (Pop III) have not been directly observed, they are thought to form from metal-free gas 
bound to dark matter minihalos since redshift $z \sim 20$ and are massive, having a
large (tens to  a few hundreds M$_{\odot}$) characteristic mass \citep[e.g.][]{Abel02, Bromm02, OShea07a,
  Turk09, Greif12_P3Cluster, Susa14, Hirano15}. Due to their high mass, they have short
lifetimes \citep{Schaerer02} and may go supernovae (SNe), and enrich their
surrounding intergalactic medium (IGM). Once the metallicity passes 
some critical metallicity, $\sim$ 10$^{-6}$ Z$_{\odot}$ if dust 
cooling is efficient \citep{Omukai05,Schneider06, Clark08} or $\sim$ 10$^{-3.5}$
Z$_{\odot}$ otherwise \citep{Bromm01, Smith09}, the gas can cool
rapidly and lower its Jeans mass and form stars in clusters.  These metal-enriched stars 
have a lower characteristic mass scale and most likely have
an initial mass function (IMF) that resembles the present-day one.

Locally, the transition from Pop III to metal enriched star formation is solely
dependent on the metal enrichment of future star forming halos from
nearby Pop III and metal enriched star SN remnants. But globally Pop III star formation 
is also regulated by radiation feedback from all UV sources \citep{Ahn12}. Lyman-Werner photons
from earlier formed stars photodissociate \hh~by the Solomon process and suppress the
formation of Pop III stars in low-mass halos \citep{Machacek01, Wise07_UVB,
  OShea08, Xu13}, delaying Pop III star formation to occur in more massive halos \citep{OShea08, Xu13}. Metal enrichment is a local process
\citep{Whalen08, Muratov12, Wise12b}, involving complex
interactions between SNe blastwaves, the intergalactic medium (IGM),
halo mergers, and cosmological accretion. As a consequence of combining these two processes,
Pop III stars may continue to form to low redshifts. This topic has been
extensively studied with semi-analytic models \citep{Scannapieco03,
  Yoshida04, Tumlinson06, Salvadori07, Komiya10}, post-processing of
numerical simulations \citep{Karlsson08, Trenti09}, and direct
numerical simulations \citep{Tornatore07, Ricotti08, Maio10_Pop32,
  Wise12a, Muratov12}. For example, \citet{Trenti09} suggested that
Pop III stars may still form at the late epoch of $z = 6$ in the under
dense regions of the universe by post-processing of cosmological
simulations with blast wave models. \citet{Muratov12} also showed that
Pop III stars continue to form until $z = 6$ using direct cosmological
simulations. But due to Pop III stars' short lifetimes \citep{Schaerer02} and high metal
 enrichment effects \citep{Nomoto06, Heger02}, the Pop III phase
of a given galaxy should be very short ($\sim$ 10 Myr), making Pop III galaxies rare at any given
redshift.  

Recently \cite{Sobral15} reported spectroscopic observations of the
strong Lyman-$\alpha$ emitter CR7 at $z=6.6$ near the end of
reionization. They interpreted the blue component and its associated
nebular emission of CR7 as a massive Pop III star cluster with a
nearby redder and more massive galaxy. This stimulated us to look for such objects in our {\em Renaissance Simulations} data \citep{Xu13,OShea15}.
We report on the results of the formation of Pop III stars at low redshifts from the low density ``Void" region of the \textit{Renaissance 
Simulations}  at $z=7.6$. Within the survey volume of $\sim$ 220 comoving Mpc$^3$, we found
a total of 14 galaxies with young Pop III stars (formed within the past $\sim$ 10 Myr, and most of which  
are still alive at $z=7.6$). We first describe the model and simulation in Section 2. In Section 3, we present our results. We show the distribution of these Pop III galaxies, their properties and UV luminosities, and their star formation
environment, including metallicity, Lyman-Werner background, and neighbor star forming galaxies. We identify
a special case of a Pop III galaxy with a nearby large metal enriched star forming galaxy and discuss its similarities and differences to CR7. 
Finally, in Section 4 we examine why Pop III stars continue to form in our simulations at such low redshifts and discuss the implications of this work.

\section{The Renaissance Simulations}
\label{sec:methods}

The \textit{Renaissance Simulations} were carried out using the
\enzo\footnote{\url{http://enzo-project.org/}} code
\citep{Bryan13}, which is an open source adaptive mesh
refinement (AMR) code that has been extensively used for simulating
cosmological structures, and in particular high-redshift structure
formation \citep[e.g.,][]{Abel02,OShea07a,Turk09,Wise12b,Wise12a,Xu13,Wise14}.
Notably, UV ionizing radiation from Pop III and metal-enriched stellar populations is followed 
using the \moray\ radiation transport solver \citep{Wise11}.
Lyman-Werner radiation from stellar sources is first calculated in the
optically-thin limit.
We then include the attenuation of LW photons
by a grey opacity approximation, the ``picket-fence" modulation factor, as a 
function of the comoving distance between the source and the observer \citep{Ahn09}. 
The properties of baryons (hydrogen, helium, and metals from stellar feedback)
are calculated using a 9-species primordial non-equilibrium chemistry
and cooling network \citep{Abel97}, supplemented by metal-dependent
cooling tables \citep{Smith09}.  Prescriptions for
Population III and metal-enriched star formation and feedback are
employed, using the same density and metallicity criteria ($< 10^{-4} \Zs$) as
\citet{Wise08_gal} and \citet{Wise12b}. Pop III stars are formed with a range of masses using a Kroupa-like
IMF but with a characteristic mass of $40 \Ms$. For each Pop III star formed its UV hydrogen ionizing and LW photon luminosities and 
lifetimes are determined by the mass-dependent model 
from \citet{Schaerer02}, and their fates and chemical yields are as described in \cite{Wise12b}.

We simulated a region of the universe that is 28.4 Mpc/h on a side (or 40 Mpc with
$h=0.71$) using the WMAP7 $\Lambda$CDM+SZ+LENS best fit cosmology \citep{Komatsu11}:
$\Omega_{M}=0.266$, $\Omega_{\Lambda} = 0.734$, $\Omega_{b}=0.0449$,
$h=0.71$, $\sigma_{8}=0.81$, and $n=0.963$.  Initial conditions were
generated at $z=99$ using MUSIC \citep{Hahn11_MUSIC}, and a
low-resolution ($512^3$ root grid) simulation was run to $z=6$ to find
regions suitable for re-simulation.  The simulation volume was then
smoothed on a physical scale of 8 comoving Mpc, and regions of high
($\langle\delta\rangle \equiv \langle\rho\rangle/(\Omega_M \rho_C) -1
\simeq 0.68$), average ($\langle\delta\rangle \simeq 0.09$), and low
($\langle\delta\rangle \simeq -0.26$) mean density (at $z=15, 12.5,
\mathrm{and}~8$, respectively) were chosen for re-simulation.  These
subvolumes, hereafter designated the ``\textit{Rare peak,}''
``\textit{Normal},'' and ``\textit{Void}'' regions, with comoving
volumes of 133.6, 220.5, and 220.5 Mpc$^3$, were resimulated with
three additional static nested grids, resulting in an effective
initial resolution of $4096^3$ grid cells and particles in the region
of interest, translating to a dark matter mass resolution of $2.9
\times 10^4$~M$_\odot$ in the same region.  We allowed further
refinement in the Lagrangian volume of the finest nested grid based on
baryon or dark matter overdensity for up to 12 total levels of
refinement, corresponding to a comoving resolution of 19 pc in the finest cells.
For more details about the calculations and scientific results about the 
{\em Renaissance Simulations}, see \citet{Xu13, Xu14}, \citet{Chen14},
and \citet{OShea15}. 

In this work, we focus on the Void simulation, which has the slowest
star formation rates and has been evolved down to $z=7.6$ to study the
late Pop III star formation at redshifts below 8. We do not include a
uniform metagalactic LW radiation background in this specific
simulation, as the LW radiation from local sources dominates the 
photon budget in our refined region.  This simulation was performed on the NCSA Blue Waters
supercomputer. Data were processed and analyzed using
\yt\footnote{\url{http://yt-project.org}} \citep{yt_full_paper} on
Blue Waters, the Gordon system at SDSC, and the Wrangler system at
TACC.

\section{Results}
\label{sec:results}

%\begin{landscape}
%\setlength{\tabcolsep}{2pt}
\begin{table*}[h]
\centering
\caption{Pop III Galaxies at $z=7.6$} 
\label{tab1}
{\scriptsize
%  \begin{tabular*}{\textwidth}{ccccccccccc}
%  \begin{tabular}{|c|c|c|c|c|c|c|c|c|c|c|c|c|}
  \begin{tabular*}{0.99\textwidth}{@{\extracolsep{\fill}} ccccccccccccc}
    \hline
    %\multicolumn{3}{}{}  & \multicolumn{3}{c}{N$_{halo}$} & & & & \multicolumn{2}{c}{f$_H$} \\
    %\cline{4-6} \cline{10-11} 
    %\noalign{\vskip 1pt}
    Halo ID & R$_{\rm vir}$(kpc) & M$_{\rm gas}$(M$_\odot$) & M$_{\rm DM}$(M$_\odot$) & M$_{\rm vir}$(M$_\odot$) & Z(Z$_\odot$,MW)  & J$_{\rm LW}$(J$_{21}$, MW) & Z(Z$_\odot$,VW) & J$_{\rm LW}$(J$_{21}$,VW) & N$_{\rm old}$ & M$_{\rm old}$(M$_\odot$) & N$_{\rm young}$ & M$_{\rm young}$(M$_\odot$) \\
    (1)      & (2) & (3)  & (4) & (5) & (6)  & (7) & (8) & (9) & (10) & (11) & (12) & (13) \\  \hline
144  & 1.72 & 2.16e7 & 8.55e7 & 1.07e8 & 1.81e-2 & 2.43  & 1.86e-1  & 1.91 & 14 & 788 & 1 & 8.86   \\
359  & 1.46 & 1.05e7 & 5.73e7 & 6.78e7 & 3.74e-2 & 0.691  &  7.07e-2 & 0.226 & 3  &  67.2 &  4 & 61.1  \\ 
448  & 1.45 & 1.22e7 & 5.53e7 & 6.75e7 & 2.27e-4 & 62.1 & 1.25e-3  & 7.79 & 0 &  0.0 & 18 & 1.15e3 \\ 
528  & 1.27 & 8.82e6 & 3.71e7 & 4.59e7 & 7.52e-5 & 11.1 & 1.91e-4  & 11.3 & 0 & 0.0 & 1  & 181    \\
599  & 1.36 & 8.50e6 & 4.60e7 & 5.45e7 & 2.95e-3 & 18.4 & 9.14e-4  & 2.46 & 11 & 616 & 1 & 15.0  \\
728  & 1.20 & 1.27e5 & 3.65e7 & 3.66e7 & 5.48e-1 & 2.81  & 8.48     & 1.39 & 31 & 1.68e3 & 4 & 28.8     \\
740  & 1.17 & 2.87e6 & 3.09e7 & 3.37e7 & 1.913    & 2.02  & 15.6    & 1.42 & 5 & 192 & 2 & 52.4     \\
763  & 1.23 & 5.77e6 & 3.38e7 & 3.96e7 & 3.76e-18 & 31.5  & 4.29e-18 & 1.75 & 0 & 0.0 & 6 & 299    \\
823  & 1.19 & 2.25e6 & 3.34e7 & 3.56e7 & 2.41e-2  & 0.282 & 2.16     & 0.450 & 14 & 358 & 2 & 17.0       \\
927  & 1.20 & 1.08e7 & 2.75e7 & 3.83e7 & 9.13e-05 & 32.7 & 1.85e-4  & 4.13 & 0 & 0.0 & 7 & 335     \\
998  & 1.10 & 5.58e6 & 2.38e7 & 2.94e7 & 8.89e-3  & 27.1 & 6.98e-4  & 3.47 & 5 & 426 & 9 & 357 \\
1200 & 1.08 & 4.47e6 & 2.23e7 & 2.68e7 & 1.51e-4 & 12.3 & 3.86e-7  & 0.554  & 5 & 242 & 0 & 0.0      \\
1316 & 1.11 & 7.76e6 & 2.20e7 & 2.98e7 & 2.77e-4 & 28.6 & 4.28e-4  & 3.38 & 4 & 237 & 5 & 285 \\
1344 & 1.07 & 3.30e6 & 2.31e7 & 2.64e7 & 8.05e-2 & 3.70  & 1.46     & 2.12 & 9 & 399 & 0 & 0.0\\
\hline
\end{tabular*}}

\parbox[t]{0.99\textwidth}{\textit{Notes:} Column (1): Halo ID. Column
  (2): Virial radius in proper kpc. Column (3): Baryon mass of the
  galaxy. Column (4): Dark matter mass of the galaxy. Column (5):
  total mass of the galaxy. Column (6): Mass-weighted
  metallicity. Column (7): Mass-weighted Lyman-Werner
  intensity. Column (8): Volume-weighted metallicity. Column (9):
  Volume-weighted Lyman-Werner intensity. Column (10): Number of Pop
  III remnants. Column (11): Mass of Pop III remnants. Column (12):
  Number of active Pop III stars. Column (13): Mass of active Pop III
  stars.}

\end{table*}
%\end{landscape}

%\begin{landscape}
%\setlength{\tabcolsep}{2pt}
\begin{table*}
\centering
\caption{Luminosities of Pop III Galaxies and its neighbors at $z=7.6$} 
\label{tab3}
{\scriptsize
%  \begin{tabular*}{\textwidth}{ccccccccccc}
%   \begin{tabular}{|c|c|c|c|c|c|c|c|}
  \begin{tabular*}{0.7\textwidth}{@{\extracolsep{\fill}} cccccccc}
    \hline
    %\multicolumn{3}{}{}  & \multicolumn{3}{c}{N$_{halo}$} & & & & \multicolumn{2}{c}{f$_H$} \\
    %\cline{4-6} \cline{10-11} 
    \noalign{\vskip 1pt} 
    Halo ID & L$_{\rm UV,Pop III}$(erg/s) & L$_{\rm UV,enriched}$(erg/s)  & Neighbor ID &M$_{1600}$ & M$_{\star}$(M$_\odot$) & L$_{\rm UV,neighbor}$(erg/s)  & l(kpc) \\
    (1)      & (2) & (3)  & (4) & (5) & (6)  & (7) & (8) \\  \hline
    \noalign{\vskip 1pt}
144 & 1.11e37 & 1.93e39  & 59   & -11.3 &  1.02e5 & 9.31e39 & 7.91 \\ 
359 & 3.65e38 & 4.08e38  & 555  &  -2.74 &  1.66e3  & 0        & 17.8 \\
448 & 5.58e40 & 0  & 618  &  -2.23 &   9.56e2 & 0        & 11.0 \\
528 & 1.24e40 & 0  & 1    & -18.0 &  6.65e7  & 7.21e42 & 11.1 \\
599 & 7.55e37 & 9.03e39  & 192   & -10.1  & 1.54e5  & 0        & 17.2 \\ 
728 & 2.33e37 & 5.37e39  & 385  & -5.58   &  2.40e3 & 0        & 10.4 \\
740 & 7.05e38 & 3.72e39  &  187 & -4.48  &  8.49e3 & 0        & 13.8 \\
763 & 9.88e39 & 0   & 238  & \ldots & 1.09e4 & 0        & 85.6 \\
823 & 1.70e37 & 3.03e39  & 1408   & -2.69  & 1.56e3  & 0        & 19.8 \\ 
927 & 1.33e40 & 0  & 10   &  -15.2  &  2.36e7 & 3.62e41 & 7.83  \\
998 & 9.85e39 & 1.44e39  & 280  &  -8.60  &  2.37e4 & 1.56e39 & 13.2\\ 
1200 & 0        & 8.06e38  & 923  & -3.19   &  1.68e3 & 0        & 14.6 \\
1316 & 1.17e40 & 0  & 50   &  -15.04 &   1.40e7 & 2.61e41 & 15.5 \\
1344 & 0        & 5.68e39  & 513  & -2.47  &  7.76e2 & 0        & 11.7 \\ \hline
\end{tabular*}
}

\parbox[t]{0.7\textwidth}{\textit{Notes:} Column (1): Halo ID. Column
  (2): Ionization UV luminosity of Pop III stars. Column (3): Ionizing
  UV luminosity of metal enriched stars. Column (4): Halo ID of nearest
  neighbor. Columns (5): Magnitude of UV in 1600 angstrom. Column (6):
  Stellar (metal enriched) mass of nearest neighbor. Column (7): Ionization UV
  luminosity of nearest neighbor.  Column (8): Distance to nearest
  galaxy in proper kpc. UV luminosities in Columns 2, 3 and 7 are
  calculated using the same formulae as implemented in the simulations
  as described in \citet{Wise12a}. M$_{1600}$ in Column 5 as
  calculated using spectrum synthesis as described in
  \citet{OShea15}.}

\end{table*}
%\end{landscape}

%\begin{landscape}
\setlength{\tabcolsep}{2pt}
\begin{table*}
\caption{Progenitors of Pop III Galaxies at $z=7.7$} 
\label{tab2}
{\scriptsize
%  \begin{tabular*}{\textwidth}{ccccccccccc}
%  \begin{tabular}{|c|c|c|c|c|c|c|c|c|c|c|c|c|c|}
  \begin{tabular*}{0.99\textwidth}{@{\extracolsep{\fill}} cccccccccccccc}
    \hline
    \noalign{\vskip 1pt} 
    ID($z=7.6$) & ID($z=7.7$) & R$_{\rm vir}$(kpc) & M$_{\rm gas}$(M$_\odot$) & M$_{\rm DM}$(M$_\odot$) & M$_{\rm vir}$(M$_\odot$) & Z(Z$_\odot$,MW)  & J$_{\rm LW}$(J$_{21}$, MW) & Z(Z$_\odot$,VW) & J$_{\rm LW}$(J$_{21}$,VW) & N$_{\rm old}$ & M$_{\rm old}$(M$_\odot$) & N$_{\rm young}$ & M$_{\rm young}$(M$_\odot$) \\
    (1)      & (2) & (3)  & (4) & (5) & (6)  & (7) & (8) & (9) & (10) & (11) & (12) & (13) & (14)\\  \hline
    \noalign{\vskip 1pt}
144 & 142  & 1.71 & 2.05e7 & 8.50e7 & 1.06e8 & 1.36e-4 & 1.08    & 1.82e-4 & 1.09 & 0 & 0 & 0 &  0     \\
359 & 355 & 1.40 & 9.87e6 & 5.36e7 & 6.34e7 & 1.88e-13 & 8.47e-2 &  1.62e-12 & 8.47e-2  & 0  &  0 &  0 &  0  \\
448 & 434 & 1.41 & 1.15e7 & 5.26e7 & 6.40e7 & 2.15e-4  & 9.43e-2 & 1.50e-3 & 9.42e-2  & 0 &  0 & 0 & 0  \\
528 & 528 & 1.27 & 8.68e6 & 3.78e7 & 4.65e7 & 4.35e-5  & 7.49    & 1.06e-4 & 7.55 & 0 & 0 & 0  & 0  \\
599 & 612 & 1.32 & 8.24e6 & 4.45e7 & 5.27e7 & 4.10e-18 & 23.7    & 4.49e-18 & 1.13 & 5 & 349 & 4 & 172  \\
728 & 740 & 1.24 & 5.71e6 & 3.64e7 & 4.21e7 & 6.49e-2  & 65.9    & 1.75e-2 & 20.3  & 3 & 178 & 32 & 2.10e3  \\
740 & 796 & 1.18 & 5.82e6 & 3.08e7 & 3.66e7 & 1.27e-4  & 0.164   & 3.50e-3  & 0.164 & 0 & 0 & 0 & 0  \\
763 & 783  & 1.11 & 5.05e6 & 3.00e7 & 3.50e7 &3.79e-18 & 6.44e-2 & 4.24e-18 & 6.41e-2 & 0 & 0 & 0 & 0  \\
823 & 846 & 1.20 & 6.16e6 & 3.24e7 & 3.85e7 & 7.58e-3  & 28.8    & 4.59e-5 & 1.26  & 8 & 358 & 8 & 182  \\
927 & 1054 & 1.19 & 1.03e7 & 2.71e7 & 3.38e7 & 8.36e-5 & 1.67    & 1.61e-4 & 1.67 & 0 & 0 & 0 & 0 \\
998 & 976 & 1.09 & 5.56e6 & 2.34e7 & 2.89e7 & 1.82e-4  &  2.61e-1 & 4.19e-4 & 0.261 & 0 & 0 & 0 & 0  \\ 
1200 & 1207 & 1.05 & 4.37e6 & 2.18e7 & 2.62e7 & 3.78e-18 & 7.18e-2& 4.13e-18 & 7.01e-2  & 0 & 0 & 0 & 0 \\
1316 & 1272 & 1.09 & 7.46e6 & 2.14e7 & 2.89e7 & 2.70e-4 & 0.483 & 4.17e-4  & 0.481 & 0 & 0 & 0 & 0 \\ 
1344 & 1266 & 1.06 & 4.41e6 & 2.32e7 & 2.76e7 & 7.99e-4 & 50.6  & 3.08e-4 & 5.47  & 0 & 0  & 3 &  334 \\ \hline  
\end{tabular*}
}
\parbox[t]{0.99\textwidth}{\textit{Notes:} Column (1): Halo ID at
  $z=7.6$. Column(2): Halo ID at $z=7.7$. Column (3): Virial radius in
  proper kpc. Column (4): Baryon mass of the galaxy. Column (5): Dark
  matter mass of the galaxy. Columns (6): Total mass of the
  galaxy. column (7): Mass-weighted metallicity. column (8):
  Mass-weighted Lyman-Werner intensity. Column (9): Volume-weighted
  metallicity. Column (10): Volume-weighted Lyman-Werner
  intensity. Column (11): Number of Pop III remnants. Column (12):
  Mass of Pop III remnants. Column (13): Number of active Pop III
  stars. Column (14): Mass of active Pop III stars.}

\end{table*}
%\end{landscape}

\begin{figure}
\begin{center}
\mbox{\includegraphics[width=1\columnwidth]{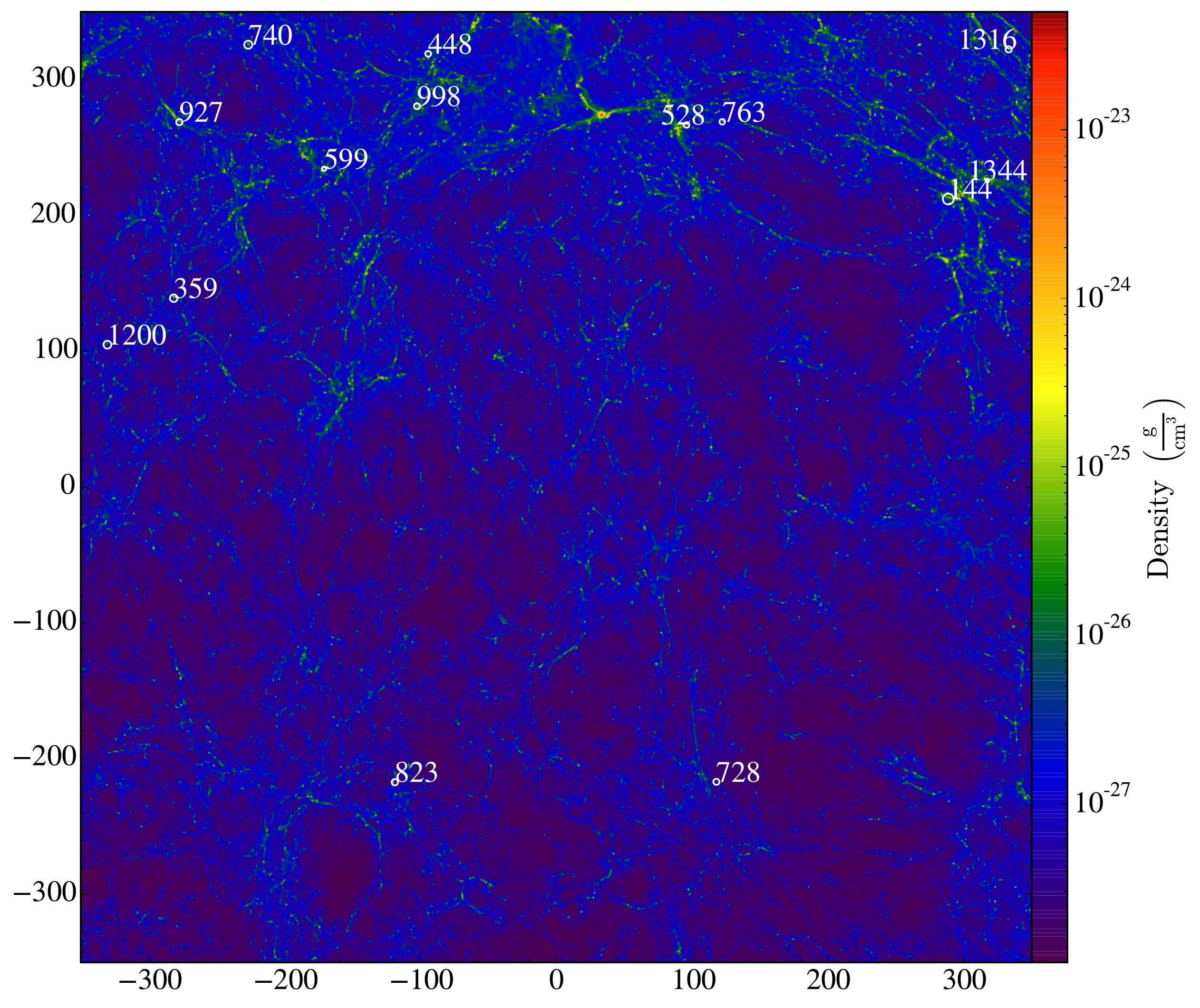}}
\mbox{\includegraphics[width=1\columnwidth]{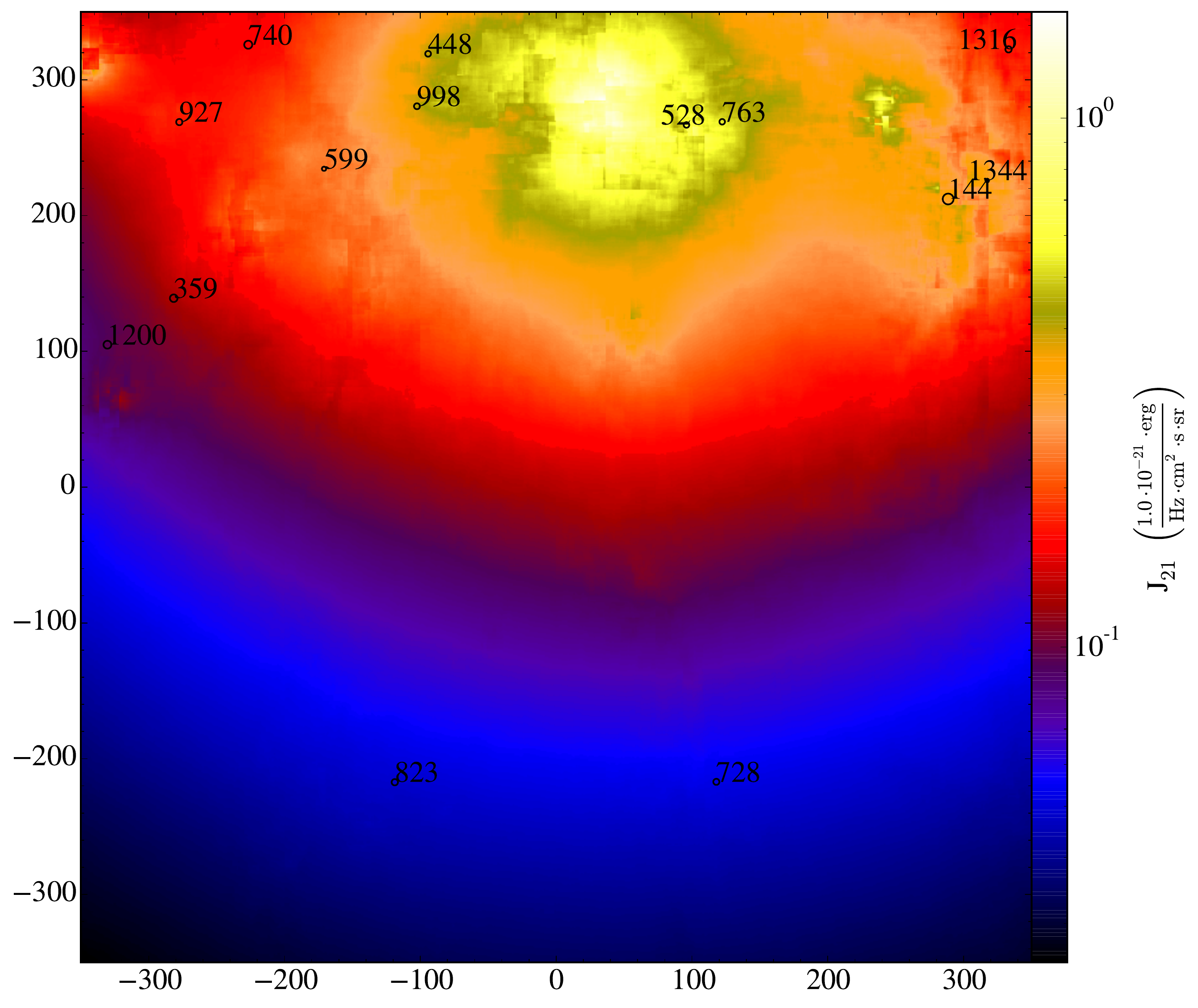}}
\mbox{\includegraphics[width=1\columnwidth]{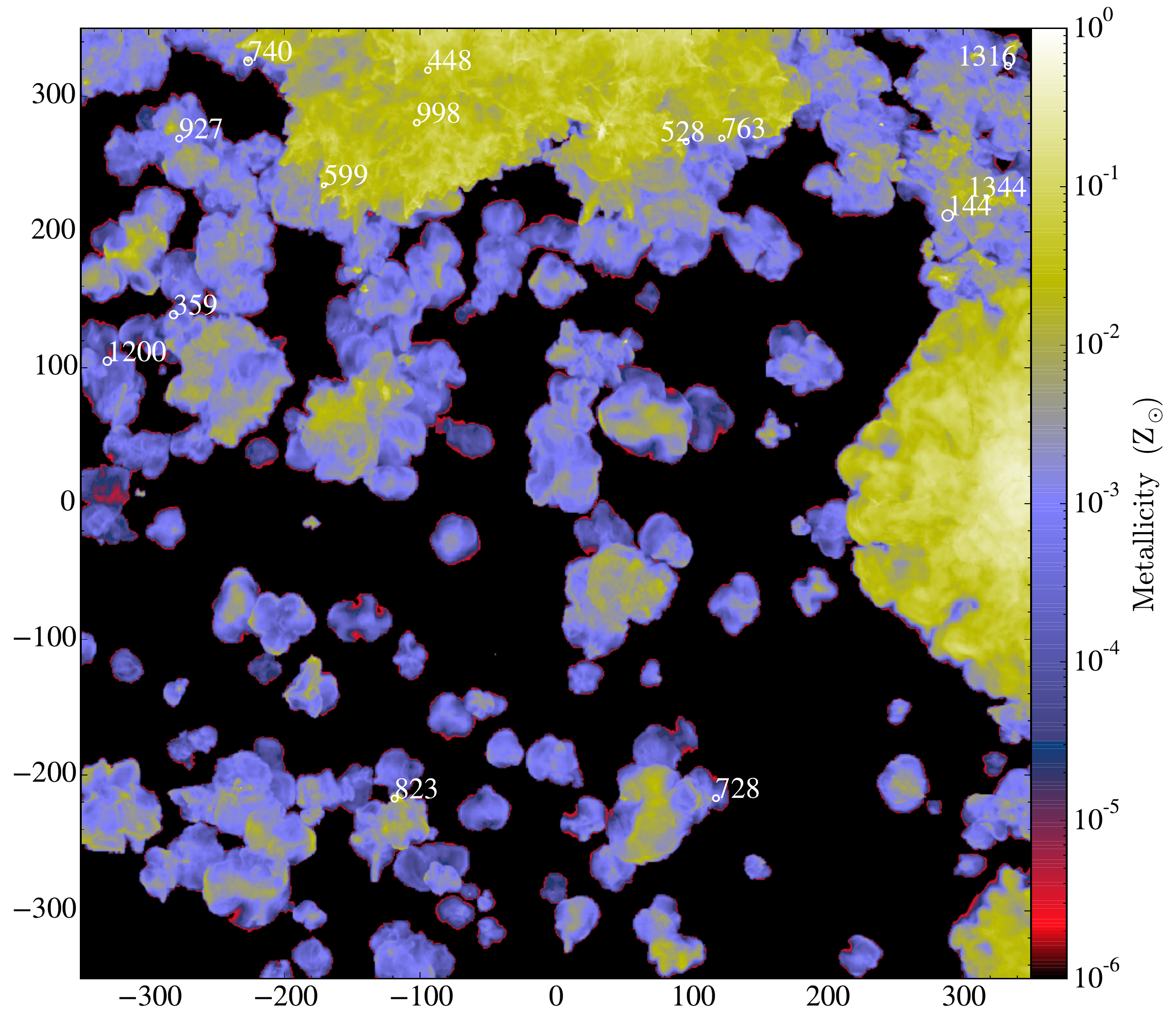}}
\end{center}
\caption{Projections through the {\em Void} survey volume of gas
  density (top), Lyman-Werner radiation intensity (middle), and
  metallicity (bottom) at $z=7.6$ with Pop III galaxies marked with
  their halo number. The survey volume is 6 comoving Mpc ($\sim$ 700
  kpc) on a side.
  \label{fig:projection}}
\end{figure}

\begin{figure}
\begin{center}
\centerline{
\includegraphics[width=1\columnwidth, clip=true]{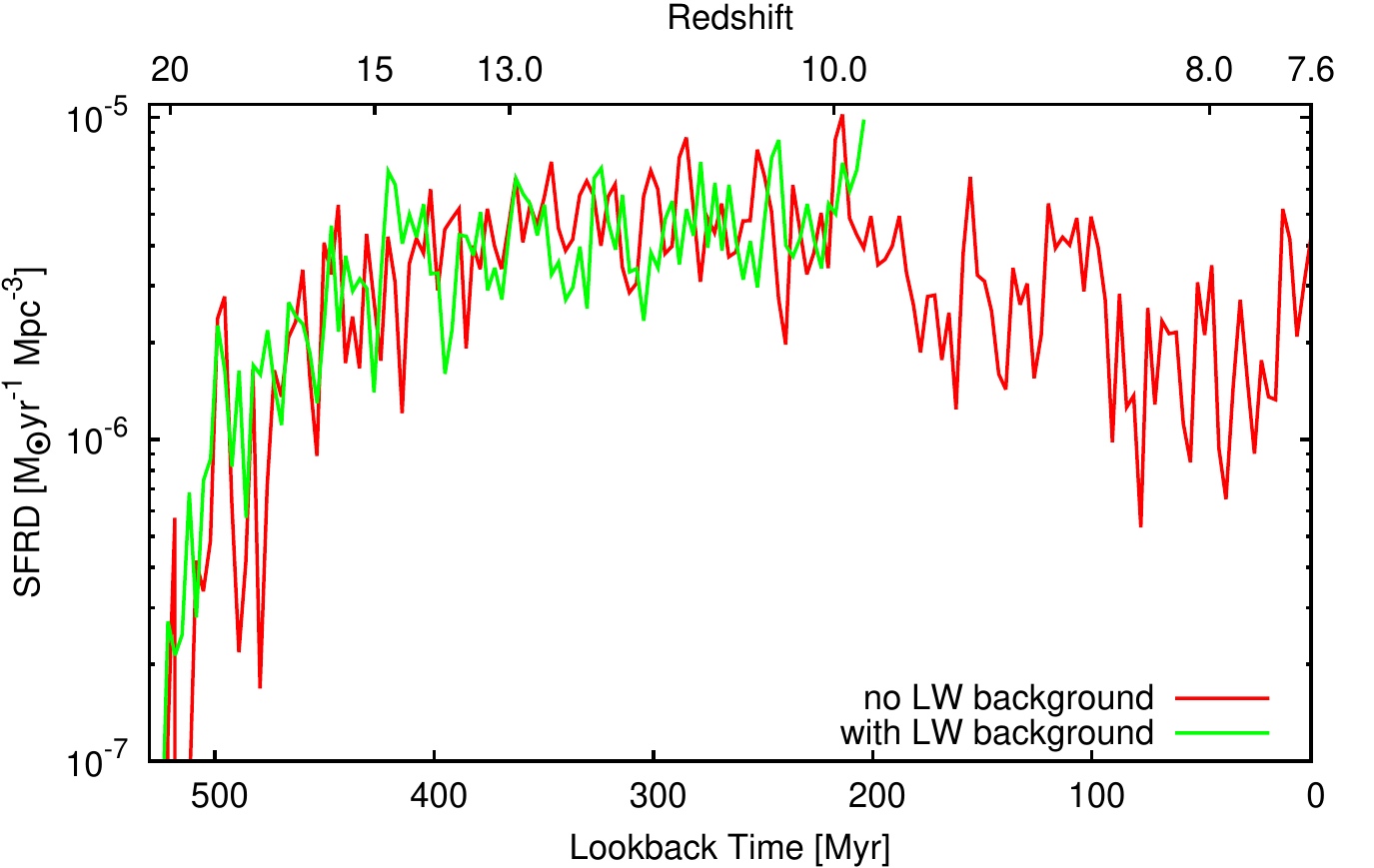}}
\centerline{
\includegraphics[width=1\columnwidth, clip=true]{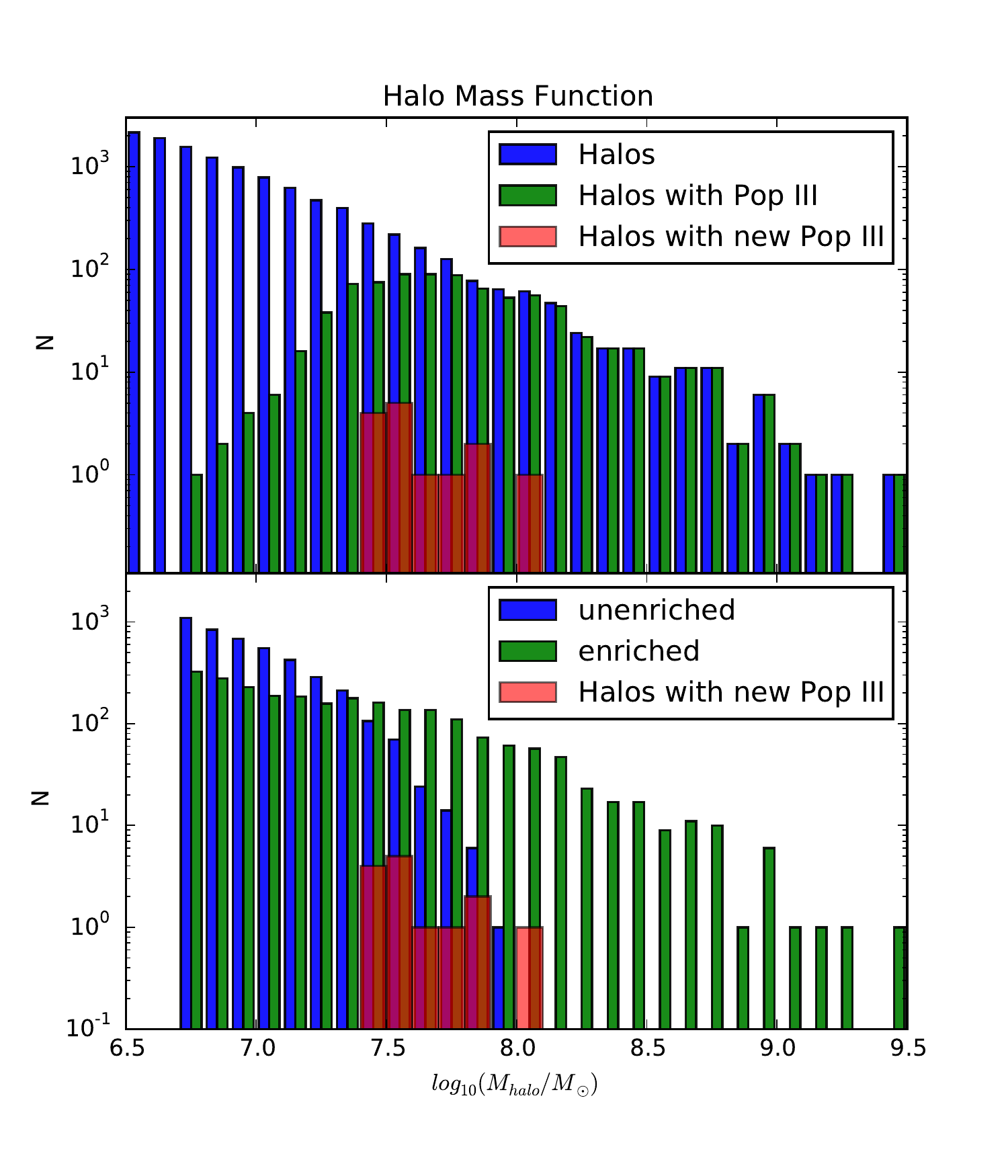}} 
\end{center}
\vspace{-1cm}
\caption{Top figure: Cosmic evolution of star formation rate density
  for Pop III. Red solid line shows the results from this simulation,
  and green dash line shows the results from the simulation of the
  same region with a uniform metagalactic Lyman-Werner background
  based on the average star formation rate density. Bottom figure:
  Halo mass function (non cumulative).  Upper panel: Blue bins show
  all halos, and green bins show halos with Pop III stars and
  remnants; lower panel: Blue bins show metal enriched halos (volume
  weighted metallicity inside virial radius [Z/H] $>$ --4), green bins
  show metal poor halos ( [Z/H] $<$ --4).  In both panels, Red bins
  shows halos with recent Pop III star formation. The upper panel
  shows all halos found by HOP halo finder, while the lower panel
  shows virialized halos only. Sub-halos (halo inside a larger halo)
  are also removed from the bottom panel.
\label{fig:hmf}}
\end{figure}

\begin{figure}
\vspace{-0.3cm}
\includegraphics[width=\columnwidth, clip=true]{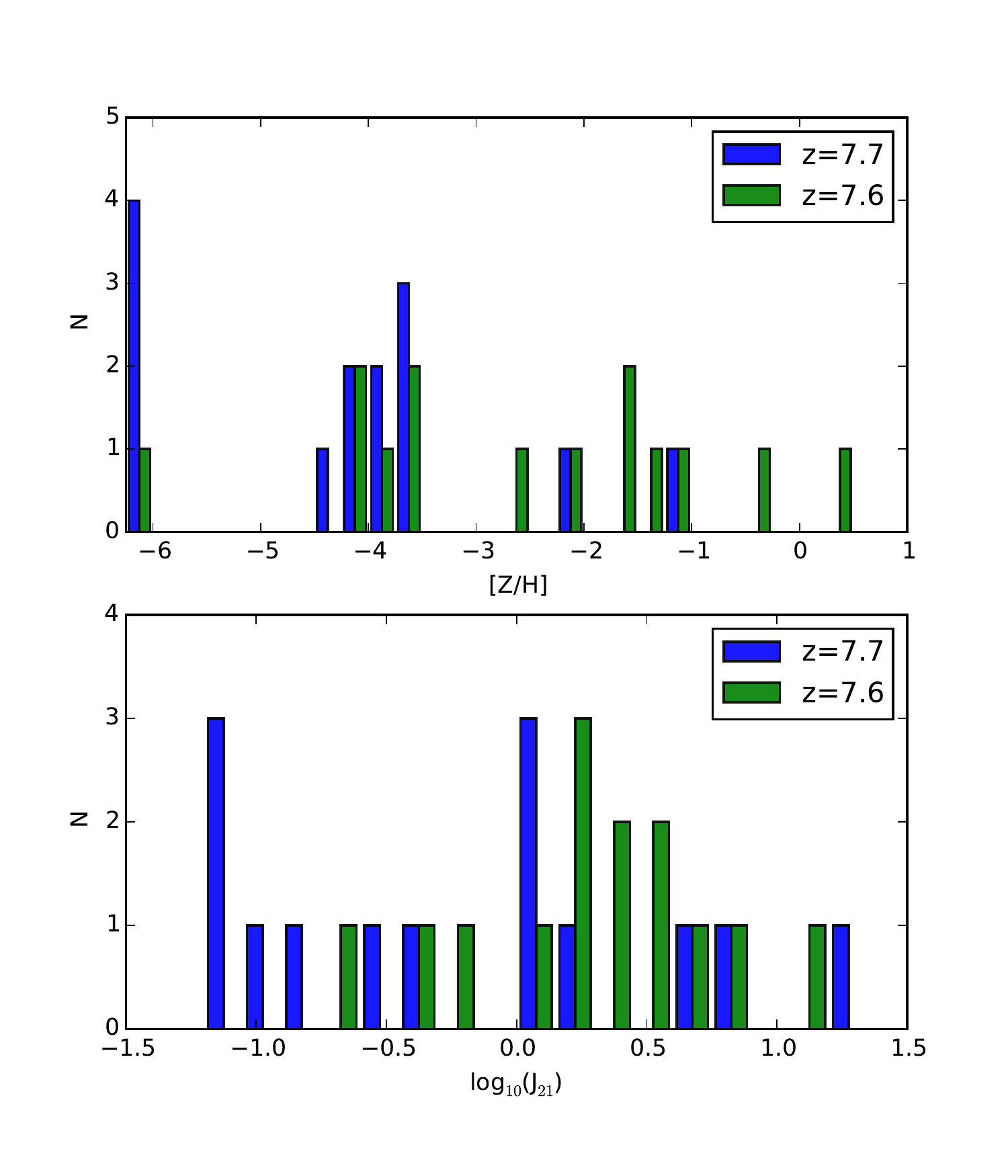}
\vspace{-0.75cm}
\caption{Mass-weighted averaged (inside virial radii) metallicity
  (top) and volume-weighted averaged Lyman-Werner intensity (bottom)
  distributions of the recent Pop III star forming galaxies. Green
  shows the halos at $z=7.6$, while blue shows their progenitors at
  $z=7.7$. The first bin in metallicity histogram shows the
  accumulcated stars with metallicity [Z/H] $<$ --6.
\label{fig:halo_distribution}}
\end{figure}

We define a Pop III galaxy as a halo containing at least one active
Pop III star, which is a star particle whose age is less than the main
sequence lifetime of a Pop III star of that mass. We treat galaxies which 
have both primordial and metal-enriched star formation going on at the same time
as Pop III galaxies. If a Pop III galaxy is a satellite of a
metal-enriched galaxy, however, we
do not consider its 
parent a Pop III galaxy. Active stars emit
ionizing and LW radiation during their lifetimes. After that they are
assumed to be non-radiating stellar remnants, which we still track
using Enzo's $N$-body machinery. After analyzing the stellar contents of
all of the halos in the survey volume, we find approximately 500 metal-enriched
galaxies in the {\em Void} simulation at $z=7.6$, whereas Pop III
galaxies are very rare with only 14 such objects in the volume.

We illustrate the distribution of the Pop III galaxies within the survey volume in Figure 1, where we show density-weighted projections 
of baryon density,  volume-weighted LW radiation intensity, and density-weighted metallicity, 
with the halo IDs of the Pop III galaxies superimposed. We list halo properties and star formation 
environments of these galaxies in three tables. Table 1 gives virial properties, metallicity, 
and LW intensity, as well as the number and mass of young (active) and old (remnant) Pop III stars in these Pop III galaxies
at $z=7.6$. As Pop III forming halos are more massive in the presence of strong LW radiation \citep{OShea08}, they are usually forming multiple Pop III stars at the same
time. Some of them have more than 10 Pop III stars with total stellar
mass of over one thousand M$_{\odot}$.
Table 2 lists for each Pop III galaxy their ionizing UV luminosities, as well as the properties of their nearest neighbor galaxies. To better show the Pop III star forming environment, we also list in Table 3 the properties of the progenitors
of these Pop III galaxies at $z=7.7$, which is about 12.1 Myr earlier. 

Pop III galaxies are located in both low and high density regions, and
have neighbors with total stellar masses ranging 
from just $10^3 \Ms$ to up to a few times 10$^7 \Ms$. At this time, Lyman-Werner
radiation, mostly from metal-enriched stars, has penetrated into the low density volume and is strong everywhere (> 0.01 J$_{21}$, J$_{21}$ is radiation intensity 
in units of 10$^{-21}$ erg s$^{-1}$ cm$^{-2}$ Hz$^{-1}$ Sr$^{-1}$),
delaying the Pop III formation and increasing the halo mass threshold
for Pop III formation to $\sim$$3 \times 10^7 \Ms$ \citep{OShea08}. By contrast, metal enrichment is local and much less efficient at affecting Pop III star formation globally \citep{Whalen08}. 
Even inside metal-enriched stellar cluster forming regions, there are still halos with pristine gas able to form new Pop III stars.

The top panel of Figure 2 shows the redshift evolution of the Pop III star formation rate density. We also show the results from the simulation of
the same region with a uniform metagalactic Lyman-Werner background based on the average star formation rate density \citep{OShea15} to z $\sim$ 10. 
LW background has little impact on the Pop III star formation rate density, since LW flux inside Pop III forming regions is dominated by local sources. 
More specifically, the LW background delays the Pop III formation for about 20 Myr for z = 13-10. At z = 9.8, there is about 7\% percent more Pop III stellar
mass in the no LW background simualtion.
The SFR density grows gradually from redshifts larger than 20 to $\sim$ 10$^{-5}$ M$_{\odot}$ yr$^{-1}$ Mpc$^{-3}$ at $z =$10 to 13.  
There is a clear trend that
the SFR density decreases slowly with decreasing redshift after $z=10$, but Pop III stars continue to form at a
significant rate at $z<$ 8.  The bottom panel of Figure 2 contains the halo mass functions showing halos with and without Pop III
stars and their enrichment status. The Pop III galaxies concentrate in
the range $M_{\rm vir}=10^{7.5}-10^8 \Ms$, as
they just pass the threshold to form Pop III in a strong LW radiation environment. 
 
Figure 3 shows the frequency of Pop III galaxies and their progenitors
as a function of metallicity and local Lyman-Werner radiation intensity. 
While Pop III galaxy progenitors have low metallicity, their halos are
quickly enriched when Pop III stars form and die as a supernova, and then 
complete the Pop III to metal-enriched star formation environment transition in less than 10 Myr. And due to their higher masses, Pop III galaxies are likely to keep
their gas after Pop III supernova explode and they begin to form metal-enriched stars immediately. Out of 14 Pop III galaxies, there are 11 galaxies also
forming metal-enriched stars. As the Lyman-Werner radiation intensity over all these halos is already at 
0.1 J$_{21}$ or more, LW radiation from  the newly formed Pop III and metal-enriched stars has only mild impacts on the total LW intensity.

\begin{figure*}
\begin{center}
\centerline{
\mbox{\includegraphics[width=\columnwidth]{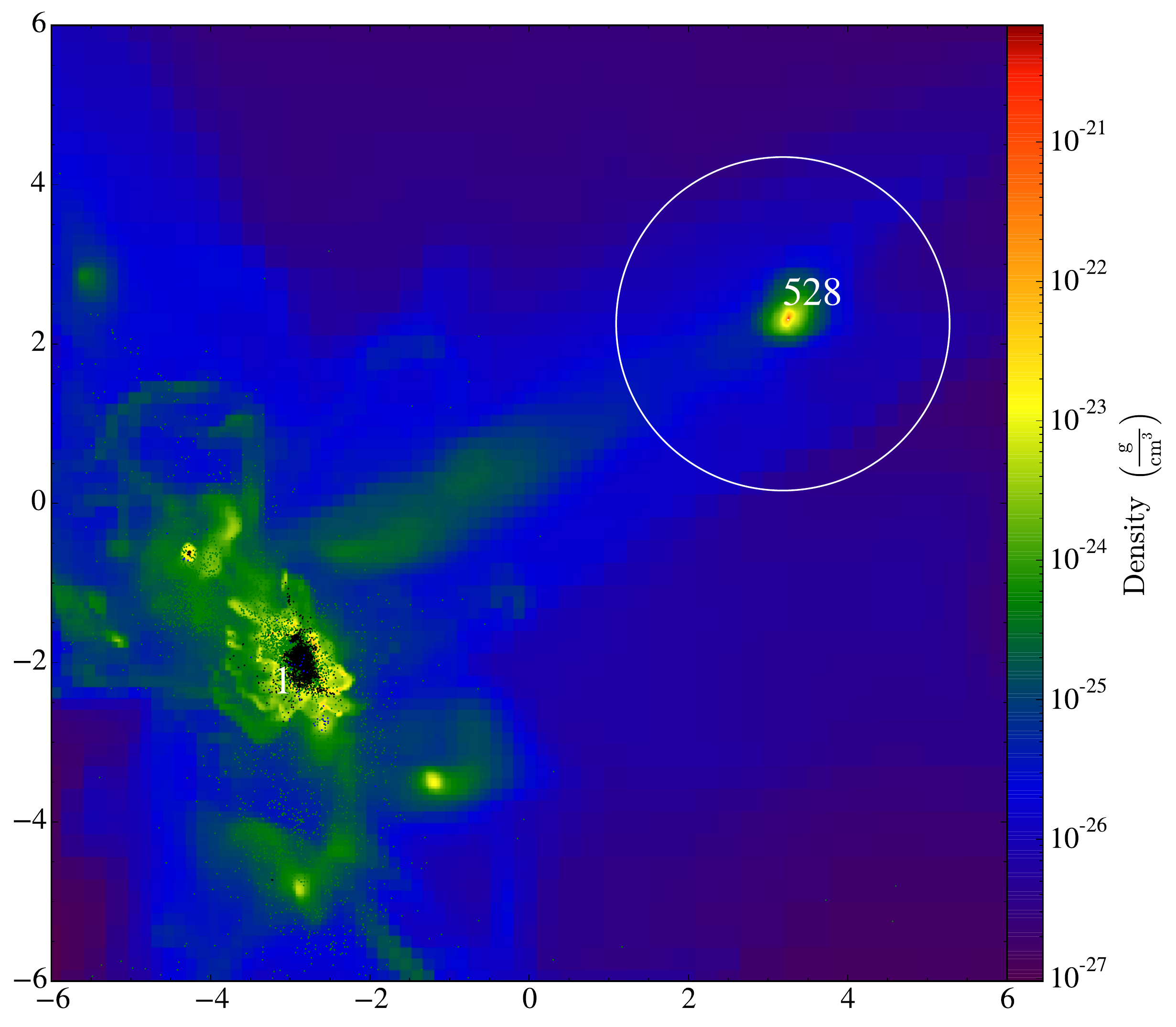}}
\mbox{\includegraphics[width=\columnwidth]{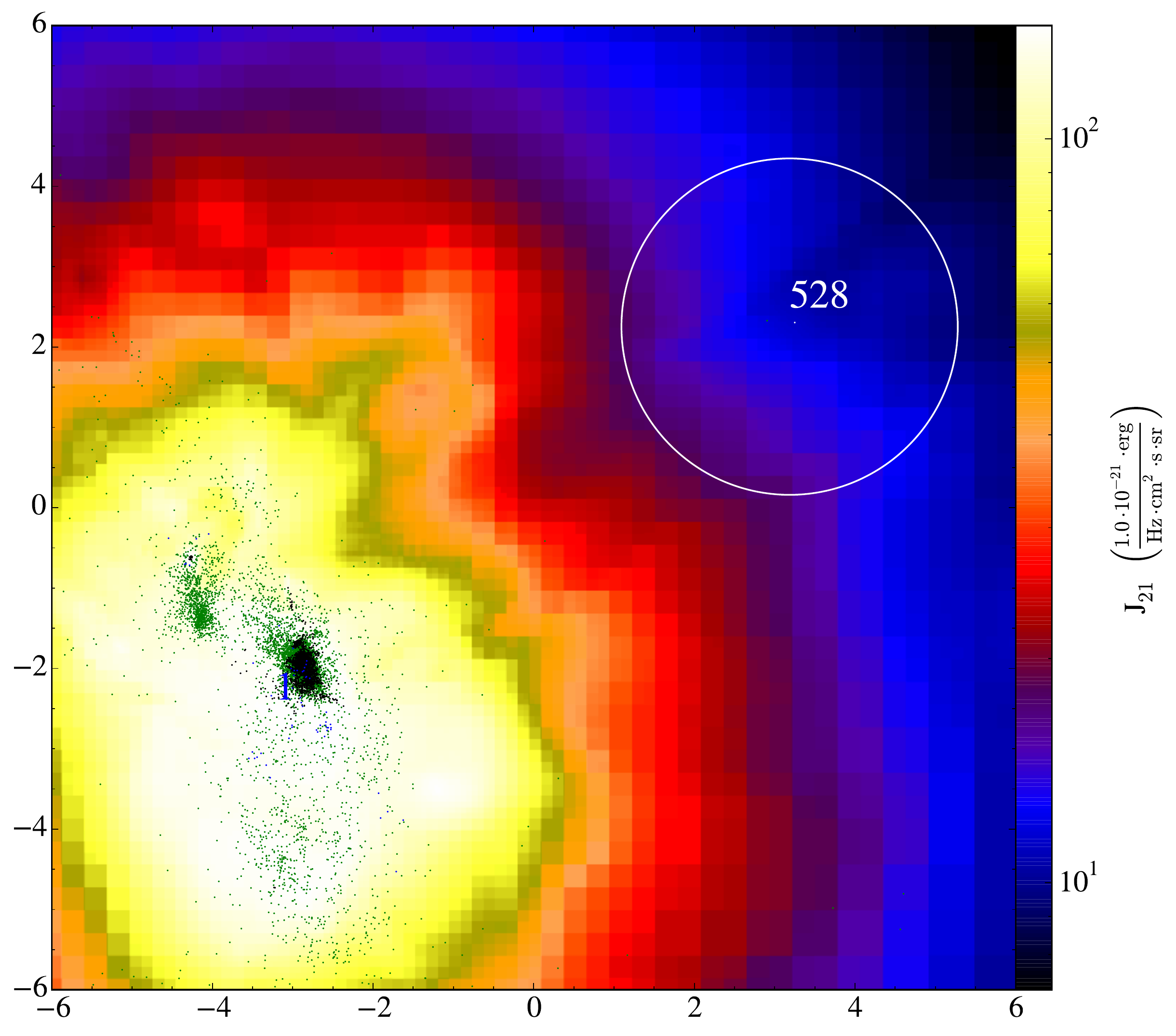}}}
\centerline{
\mbox{\includegraphics[width=\columnwidth]{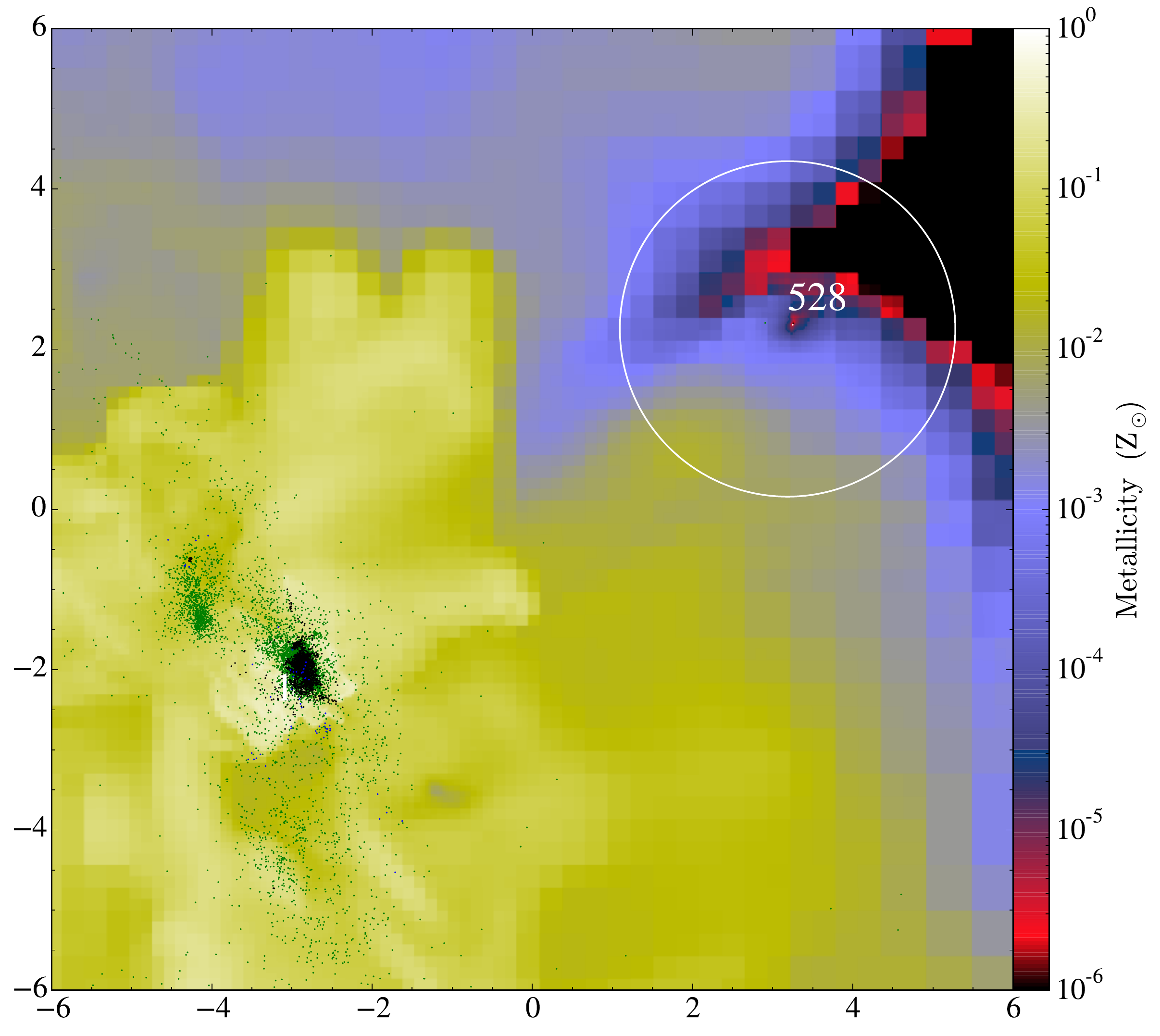}}
\mbox{\includegraphics[width=\columnwidth]{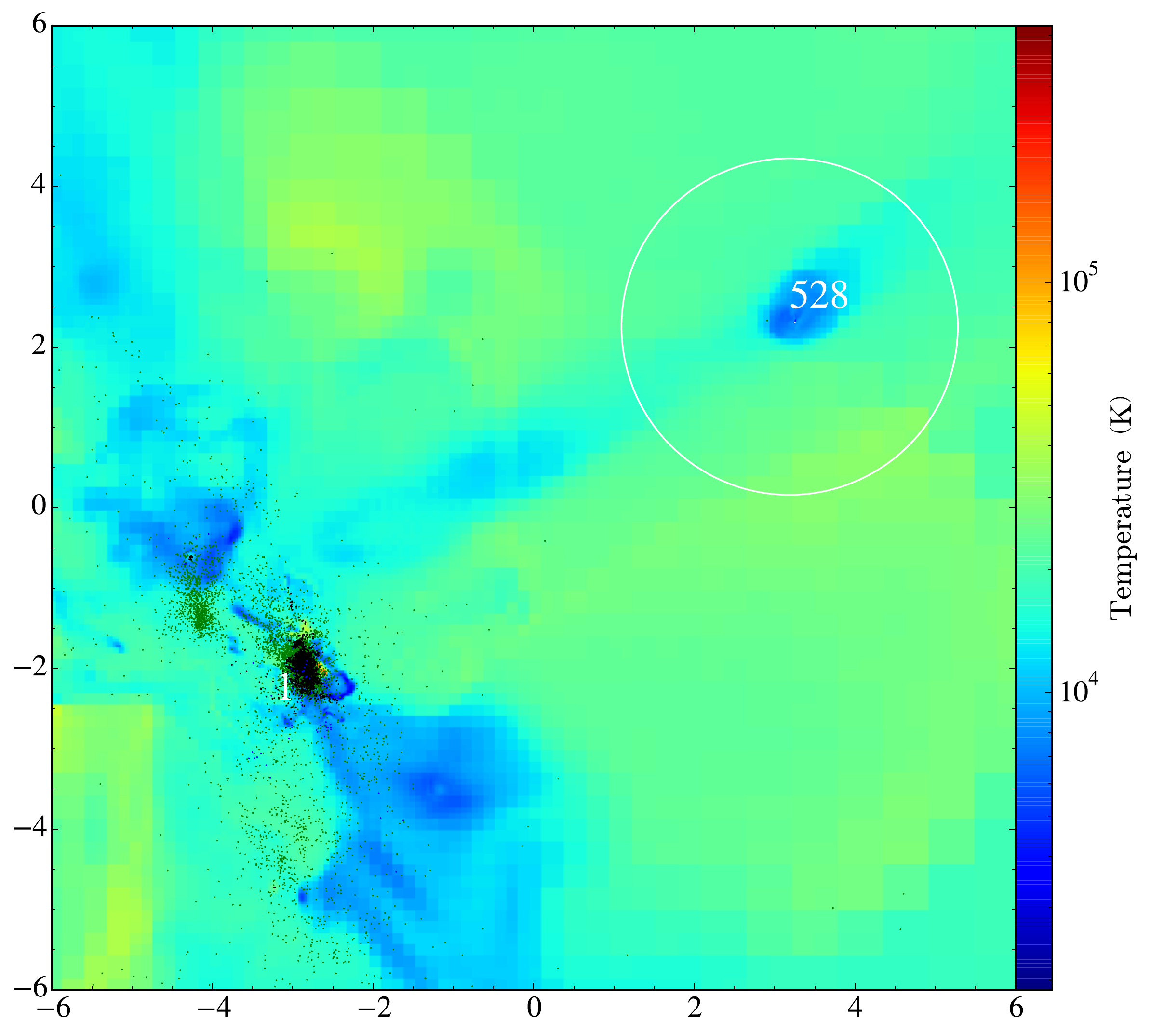}}}
\end{center}
\caption{Projections of (from the top left panel, counter-clockwise) gas density, metallicity, temperature and Lyman-Werner radiation intensity of
a volume including Halo 528 with a single young Pop III star and its metal enriched star forming galaxy neighbor. The projection volume is a cube of a side of 12 kpc.
White dot shows the active Pop III star in Halo 528. Green dots are old metal enriched star clusters, while black dots are UV active metal enriched star clusters.
\label{fig:projection_528}}
\end{figure*}

Figure \ref{fig:projection_528} shows projections of gas density,
metallicity, temperature, and Lyman-Werner flux of a metal-enriched
galaxy with a stellar mass $M_\star = 6.7 \times 10^7 \Ms$ and a
nearby halo (ID 528; $M_{\rm vir} = 4.6 \times 10^7 \Ms$; $M_{\rm gas}
= 8.8 \times 10^6 \Ms$) that is 11.1 kpc away and has only formed a
single Pop III star.  The UV luminosity of the Pop III galaxy is
only 1.2 $\times$ 10$^{40}$ erg s$^{-1}$, or about 1/600$^{\rm th}$ of
its metal-enriched galaxy neighbor. The feedback of the metal-enriched
stars from the large galaxy neighbor has heated and enriched the gas
surrounding the Pop III galaxy, but fails to change the pristine
status of its core region, allowing the Pop III star to form. As the
radiation ionizes the enriched medium surrounding the Pop III galaxy,
metal nebular emission lines will be produced even though the ionizing
source is metal-free.

We compare this simulated system with the inferred properties of CR7
that could be interpreted as a Pop III galaxy with a stellar mass
$\sim 10^7 \Ms$ with an evolved galaxy with a stellar mass of $\sim
10^{10} \Ms$ that is 5 (projected) kpc away \citep{Sobral15}.  The
total Ly$\alpha$ luminosity of CR7 was measured to be $8.5 \times
10^{43} \unit{erg s}^{-1}$, which is an order of magnitude brighter
than the UV luminosity of the simulated galaxy, $7.2 \times 10^{42}
\unit{erg s}^{-1}$ (M$_{1600} = -18.0$).

The proximity of the two systems is the only similarity between CR7
and the simulated galaxy pair.  While our simulated system is a low-mass
analog to the Pop III interpretation to CR7, it is unlikely that CR7
is powered by a Pop III-like stellar population because rapid chemical
enrichment will terminate any subsequent Pop III star formation in
Halo 528.  Such enrichment will originate externally from metal-rich
outflows, as seen in the metallicity projection of Figure
\ref{fig:projection_528}, and internally from Pop III supernovae.  For
a halo to form $\sim 10^7 \Ms$ of Pop III stars, it will need to
remain chemically pristine and form all of its stars in one burst,
avoiding internal enrichment.  This scenario requires a nearby galaxy
that provides the Lyman-Werner flux to suppress Pop III star
formation, but the halo cannot be too close as it will be externally
enriched.  Our simulated system illustrates such a ``Goldilocks''
scenario with the pristine halo having a gas mass $M_{\rm gas} = 8.8
\times 10^6 \Ms$.  We do not expect any pristine halos to have a much
higher gas mass, and thus it is nearly impossible to produce a Pop III
galaxy with a stellar mass $\sim 10^7 \Ms$ that would require all of
the gas to be converted into stars.  We conclude that the Pop III-like
interpretation of CR7 is highly unlikely \citep{Visbal16}, and its
nature could be explained by either a direct collapse black hole
\citep{Pallottini15, Agarwal15, Hartwig15, Smidt16, Smith16} or a
young metal-poor galaxy.  In the latter case, the very blue color of
CR7 could be caused by a sight-line that is aligned with a
photo-ionized and photo-heated channel in which ionizing photons are
escaping into the nearby intergalactic medium and destroying any dust
grains within this channel \citep[cf.][for a $z=3$
analog]{Vanzella16}.

%%% Local Variables: 
%%% mode: latex
%%% TeX-master: "ms"
%%% End: 

\section{Summary and Discussion}
\label{sec:conclusions}

The key results of this work are as follows:

\begin{enumerate}

\item Pop III stars continue to form at significant rates down to
  $z=7.6$, the termination redshift of the simulation. Pop III stars can still form in pristine pockets of gas inside metal enriched star forming regions.

\item Halos hosting late Pop III star formation (``Pop III galaxies'')
  are significantly more massive at lower redshift than at higher redshift, with virial masses between 5 $\times$ 10$^{7}$ and 1 $\times$ 10$^8$
M$_{\odot}$ at $z \simeq 7.5$. This is due to the strong LW radiation from metal
enriched star forming galaxies in their vicinity. Suppression of Pop
III star formation in lower mass halos ($M_{\rm vir} < 10^7 M_\odot$) extends the Pop III era. 

\item Pop III galaxies form multiple Pop III stars and also contain Pop III stellar remnants.
A single Pop III galaxy could have more than 10 active Pop III stars
and as much as $10^3 \Ms$ in metal-free stellar mass. 

\item One of our simulated Pop III and metal-enriched galaxy pairs is
  a low-mass analog of the Pop III-like interpretation of the observed
  CR7 galaxy \citep{Sobral15}.  However, we rule out this scenario as
  metal enrichment will quickly halt any further metal-free star
  formation after the initial burst.

\end{enumerate}

Late formation of Pop III stars is caused by the combined effects of slow local metal enrichment
and strong global LW radiation. While the metal-enriched bubbles have radii of tens of kpc at most, Lyman-Werner
radiation from clustered metal enriched star galaxies can travel tens of Mpc deep into low density regions. 
Lyman-Werner radiation increases the halo mass threshold for Pop III
formation \citep{OShea08}, and furthermore in these larger halos, mixing
timescales are longer for the metal-rich blastwave to advect
from the virial radius to the star forming regions \citep[as described in some detail in][]{Smith15}. Therefore, Population
III star formation is further delayed until the halo grows through
slow cosmological accretion.

The formation of Pop III stars likely continues to even lower redshifts in this simulation. In the survey volume at
$z=7.6$, only 6.22\% of the volume and 13.01\% of the gas mass is
enriched to [Z/H] $>$ --4. The entire volume with lower metallicity gas is immersed in strong LW radiation from metal enriched star forming galaxies. The fraction of the volume with Lyman-Werner intensity stronger than 1, 0.1, and 0.01 J$_{21}$, is 5.12\%, 55.37\% and 100\%, respectively.  The metal 
enrichment rate is even lower than the ionization rate. At this redshift, there is only 16.1\% of the volume or 17.7\% of the mass
of hydrogen that has been ionized. It is reasonable to expect that Pop III star formation will happen even after reionization is complete..    

The survey volume less than 300 coming Mpc$^3$ is very small, so any results from this simulation of late Pop III formation and Pop III galaxies should be considered as 
the upper limit on redshfits. Our simulation (with 0.062 Pop III galaxies per comoving Mpc$^3$ at z = 7.6) suggests that Pop III galaxies are not rare at z = 7-8, at 
least in under dense regions. And out of these simulated 14 Pop III galaxies, there is one in a Pop III -- metal enriched galaxies pair resembling CR7 in spatial distribution. 
Although the sample size is too small to draw any statistical significant conclusions, it is reasonable to expect there are many close Pop III -- metal enriched galaxies pairs
formed at this or even lower redshifts in the vast low density regions. Some of them may have Pop III stellar mass upto thousands M$_{\odot}$ as other Pop III galaxies found 
in our simulations and could be much easier to be deteched.

We do not include a uniform metagalactic Lyman-Werner background based on the average star formation rate density; we only include Lyman-Werner flux from star formation in the refined region of our simulation. However we expect that the inclusion of LW background would not change our results significantly. 
This is because any additional
LW radiation would only increase the halo mass threshold for Pop III formation and hence further delay their formation. These
effects are weak when the LW intensity is higher than J$_{21}$ $>$ 0.1 \citep{OShea08}, which has been reached by
local sources only in our present simulation.  To check this
hypothesis, we have performed the {\em Void} simulation with a LW
background derived from the {\em Normal} region simulation, stopping
at $z=10$.
We find that this simulation with an uniform LW background has
a very similar Pop III star formation history as the one reported on
in this paper, as shown in the top panel of Figure 2. 
    
The primary results of this paper -- that Pop III star formation extends to late stages of reionization, 
and that Pop III stars form in clusters in higher mass halos -- are significant for observations and our understanding of the 
formation of the first generation stars in the universe. It is beyond the sensitivity of JWST to
directly see individual Pop III stars forming at high redshifts of $z=15-20$, except possibly their death 
through PISN explosions \citep{Wise05, Trenti09, Hummel12,
  Whalen13}. However, it could be much easier to discover and study Pop III
clusters in a post-reionization environment as their formation is not halted during
the early stages of reionization.

One possible important consequence of late Pop III star formation is that they might 
become strong X-ray sources if their stellar remnants are in binary systems \citep{Xu14} with impacts on the X-ray background. X-rays photo-heat and pre-ionize
the IGM homogeneously on tens to hundreds of Mpc scales, and may be important to understand results of
Ly$\alpha$ forest spectra  \citep{tytler09} and 21-cm  signatures \citep{Ahn15}. A detailed analysis of the X-ray 
background from Pop III binaries from the very beginning of Pop III stars ($z\sim 25$) to the end of the reionization will 
be reported in a separate paper.

%%% Local Variables:
%%% mode: latex
%%% TeX-master: "ms"
%%% End:

\acknowledgements

This research is part of the Blue Waters project, which a joint effort
supported by the NSF (award number ACI-1238993) and the state of
Illinois, using NSF PRAC OCI-0832662.  This research was supported by
NSF grants PHY-0941373, AST-1109243, AST-1211626, and AST-1333360, and
NASA grants NNX12AC98G, HST-AR-13261.01-A and HST-AR-13895.001.  This
work was performed using the open-source {\sc Enzo} and {\sc yt} codes,
which are the products of collaborative efforts of many independent
scientists from institutions around the world. Their commitment to
open science has helped make this work possible.

%\newpage

%\bibliographystyle{apj}
%\bibliography{ms}

\end{document}